\documentclass[aps,pra,twocolumn,superscriptaddress]{revtex4}

\usepackage{amsmath,amsfonts,amssymb}
\usepackage{graphicx,float,calc}
\usepackage{color,bm}
\usepackage{ulem}
\usepackage{braket}
\usepackage{hyperref}
\usepackage{graphicx,epstopdf}
\usepackage{CJK}

\newcommand{\ehbar}{\hbar_{\text{eff}}}

\begin{document}
\begin{CJK*}{GBK}{song}  %%%% begin Chinese, Japanese, and
%\title{Diagnosing the $\cal{PT}$-symmetry breaking by the dynamics of the out-of-time-order correlators in a non-Hermitian kicked rotor model}
\title{Scaling laws of the out-of-time-order correlators at the transition to the spontaneous $\cal{PT}$-symmetry breaking in a Floquet system}
\author{Wen-Lei Zhao}
\email{wlzhao@jxust.edu.cn}
\affiliation{School of Science, Jiangxi University of Science and Technology, Ganzhou 341000, China}
\author{Ru-Ru Wang}
\affiliation{School of Science, Jiangxi University of Science and Technology, Ganzhou 341000, China}
\author{Han Ke}
\affiliation{School of Science, Jiangxi University of Science and Technology, Ganzhou 341000, China}
\author{Jie Liu}
\email{jliu@gscaep.ac.cn}
\affiliation{Graduate School, China Academy of Engineering Physics, Beijing 100193, China}
\affiliation{HEDPS, Center for Applied Physics and Technology, and College of Engineering, Peking University, Beijing 100871, China}

\begin{abstract}
We investigate both numerically and analytically the dynamics of out-of-time-order correlators (OTOCs) in a non-Hermitian kicked rotor model, addressing the scaling laws of the time dependence of OTOCs at the transition to the spontaneous $\mathcal{PT}$ symmetry breaking.
In the unbroken phase of $\mathcal{PT}$ symmetry, the OTOCs increase monotonically and eventually saturate with time, demonstrating the freezing of information scrambling. Just beyond the phase transition points, the OTOCs increase in the power-laws of time, with the exponent larger than two. Interestingly, the quadratic growth of OTOCs with time emerges when the system is far beyond the phase transition points. Above numerical findings have been validated by our theoretical analysis, which provides a general framework with important implications for Floquet engineering and the information scrambling in chaotic systems.
\end{abstract}
\date{\today}

\pacs{03.65.-w, 03.65. YZ, 05.45.-a, 05.45.Mt}

\maketitle

%%%%%%%%%%%%%%%%%%%%%%%%%%%%%%%%%%%%%%%%%%%%%%%%%%%%%%%%%%%%

\section{Introduction}
Non-Hermiticity has been regarded as a fundamental modification to the conventional quantum mechanics~\cite{Markum99,Stephanov99,Berry2004,Graefe2008,Rotter2009,Ott2013,Harsh2014}, a subclass of which with $\cal{PT}$ symmetry even displays the transition from the real energy spectrum to complex one. Such intrinsic spontaneous $\cal{PT}$-symmetry breaking occurs at the exceptional points (EP), at which both the eigenstates and eigenvalues coalesce~\cite{Bender1998,Bender2002,Mosta2002,Bender2007,Mosta2010,Harsh2010,Ganainy2018,Klaiman2008,Musslimani2008}.  The existence of EP leads to rich physics, such as the enhancement of precision in quantum sensors~\cite{Cai20}, the topological phase transition~\cite{Bergholtz21,ZhongWang18,Mzhao1,FYu,Mzhao2}, the nonadiabatic transition~\cite{FengYu21,WYWang22}, and the unidirectional propagation of light~\cite{YinHuang}, just to name a few. Theoretical advances have enabled exponential realizations of $\cal{PT}$ symmetric systems in various fields, such as optical settings~\cite{Sergi2014,Makris2008,Ganainy2007,Guo2009,Regensburger2012,Hodaei2014,Feng2014,Jiahua2016,Longhi2009,Longhi2010,Ruter2010,YongmeiXue}, electronic circuits~\cite{Stegmaier2021}, and optomechanical systems~\cite{Luxy15}. Moreover, the extension of Floquet-driven systems to the $\mathcal{PT}$ symmetric regime has opened up unique opportunities for understanding fundamental concepts such as quantum chaos~\cite{Bender09} and quantum-classical transition~\cite{Bender10,wlzhao23a}. Interestingly, chaos is found to facilitate the scaling law of the spontaneous $\mathcal{PT}$ symmetry breaking in a $\mathcal{PT}$ symmetric kicked rotor (PTKR) model~\cite{West2010}. This system even displays ballistic energy diffusion~\cite{Longhi2017} and the quantized acceleration of momentum current~\cite{Zhao19}, which enriches our understanding on the unique transport phenomena in the presence of chaos.

The dynamics of OTOCs, originally introduced by Lakin {\it et al.}, in the study of quasiclassical theory of superconductivity~\cite{Larkin1969}, has received extensive studies in the fields of high energy physics~\cite{Roberts2016,Maldacena2016,Polchinski2016}, condensed matter physics~\cite{Bohrdt2016,Garttner2017,Banerjee2017,Shen2017,Fan2017,Huang2017} and quantum information~\cite{Li2017,Weinstein22,Hu23}. It has been found that OTOCs can effectively detect quantum chaos~\cite{Pappalardi22,GMata18,JiaoziWang21,Kidd2021}, quantum thermalization~\cite{Balachandran21}, and information scrambling~\cite{Harris2022,Roberts2022,Zhang2019,Yan2020,Patel2017,WLZhao23}. In the semiclassical limit, the exponential growth of OTOCs is governed by the Lyapunov exponent of classical chaos, which demonstrates a route of quantum-classical correspondence~\cite{Hashimoto2017}. In Floquet-driven systems, OTOCs has been used to diagnose dynamical quantum phase transition~\cite{Zamani2022} and entanglement~\cite{Martin2018,Lewis-Swan2019}.
Intrinsically, we previously found a quantized response of OTOCs when varying the kicking potential of the PTKR model~\cite{WlZhao22}. State-of-art experimental advances have observed different kinds of OTOCs in the setting of nuclear magnetic resonance~\cite{Wei2018,Nie2019}, trapping ions~\cite{Landsman2019} and qubit under Floquet engineering~\cite{Zhao2021}.

In this context, we both numerically and analytically investigate the dynamics of OTOCs when the PTKR model is in different phases of $\cal{PT}$-symmetry. We use a machine learning method, namely a long short-term memory network (LSTM), to classify the phase diagram of $\cal{PT}$-symmetry breaking and extract the phase boundary in a wide range of system parameters. We find that in the unbroken phase of $\mathcal{PT}$ symmetry, OTOCs increase monotonically with time evolution and eventually saturate, demonstrating the freezing of operator growth. We analytically prove that the saturation of OTOCs is a power-law function of the real part of the kicking potential. In the broken phase of the $\mathcal{PT}$ symmetry, we find a power-law increase of OTOCs with time, for which the characteristic exponent is larger than two when the system is just beyond the phase transition point, and is equal to two for the system far beyond the phase transition point. Through the detailed analysis of the wavepacket's dynamics in the time reversal process, we uncover the mechanisms of both the dynamical localization and the power-law increase of OTOCs. Our investigations reveal that the dynamics of OTOCs can be utilized to diagnose spontaneous $\cal{PT}$-symmetry breaking.

The paper is organized as follows. In Sec.~\ref{Sec-PTB}, we describe the PTKR model and show the scaling-law of spontaneous $\cal{PT}$-symmetry breaking. In Sec.~\ref{OTCPT}, we show the scaling-laws of the dynamics of OTOCs at the transition to the $\cal{PT}$-symmetry breaking. Sec.~\ref{TheoAnaly} contains the theoretical analysis of the scaling-laws of OTOCs. The conclusion and discussion are presented in Sec.~\ref{Sum}.

\section{Transition to spontaneous $\cal{PT}$-symmetry breaking in Floquet systems}\label{Sec-PTB}

\subsection{Model}

The Hamiltonian of the PTKR model
in dimensionless unites reads
\begin{equation}\label{nschoreq}
\textrm{H} = \frac{p^2}{2} +V(\theta)\sum_n \delta(t-t_n)\;,
\end{equation}
where the kicking potential $V(\theta)=K\left[ \cos(\theta) + i \lambda \sin(\theta)\right]$ satisfies the $\cal{PT}$-symmetric condition $V(\theta)=V^{*}(-\theta)$~\cite{Longhi2017,West2010,Zhao19}. The parameters $K$ and $\lambda$ indicate the strength of the real and imaginary parts of the kick potential, respectively. The $p=-i\ehbar {\partial}/{\partial \theta}$ is angular momentum operator, $\theta$ is the angle coordinate, and $\ehbar$ denotes the effective planck constant. The time $t_n (=0,1,2\ldots)$ is integer, indicating kicking numbers. The eigenequation of angular momentum operator is $p |\varphi_n \rangle = n \ehbar |\varphi_n \rangle$ with eigenstate $\langle \theta |\varphi_n \rangle= e^{i n \theta}/\sqrt{2\pi}$ and eigenvalue $p_n= n \ehbar$. On the basis of $|\varphi_n \rangle$, an arbitrary quantum state can be expanded as
$|\psi \rangle = \sum_n \psi_n |\varphi_n \rangle$.

For time-periodic systems, i.e., $\textrm{H}(t+T)=\textrm{H}(t)$, the Floquet theory predicts the eigenequation of the evolution operator $U|\psi_{\varepsilon}\rangle =  e^{-i \varepsilon}|\psi_{\varepsilon}\rangle$, where the eigenphase $\varepsilon$ is referred to as quasienergy. One-period time evolution of a quantum state of the PTKR system is given by $|\psi(t_{j+1}\rangle =U|\psi(t_{j}\rangle $ with the Floquet operator \begin{equation}\label{evol}
U = U_fU_K=\exp\left(-\frac{i}{\ehbar} \frac{p^2}{2}\right) \exp\left[-\frac{i}{\ehbar}V(\theta)\right]\;.
\end{equation}
This demonstrates that in numerical simulations, one period evolution is split into two steps, namely the kicking evolution $U_K$ and the free evolution $U_f$. The kick evolution is realized in angle coordinate space, i.e., $\psi'(\theta)=U_K(\theta) \psi(\theta,t_j)$. Then, one can utilize the fast Fourier transform to change the state $\psi'(\theta)$ to angular momentum space, thereby obtaining its component $\psi'_n$ on the eigenstate $|n\rangle$. Finally, the free evolution is conducted in angular momentum space, i.e., $\psi_n(t_{j+1})=U_f(p_n)\psi'_n$. By repeating the same procedure, one can get the quantum state at arbitrary time~\cite{Casati79}.

\subsection{Spontaneous $\cal{PT}$-symmetry breaking}

It is straightforward to prove that the Floquet operator of the PTKR satisfies the $\cal{PT}$ symmetry $U =  (\mathcal{P}\mathcal{T})^{\dagger}U\mathcal{P}\mathcal{T}$, where $\cal{P}$ and $\cal{T}$ are the parity and time reversal operators, respectively. Based on conventional understanding of quantum mechanics, one asserts that the two operators, i.e., $U$ and $\cal{PT}$ have simultaneous eigenstates, that is to say, the quasieigenstate $|\psi_{\varepsilon}\rangle$ is also the eigenstate of the $\cal{PT}$ operator, i.e., $\cal{PT}|\psi_{\varepsilon}\rangle=\pm |\psi_{\varepsilon}\rangle$. This conclusion is indeed valid for positive quasienergies $\varepsilon>0$. However, a notable feature of the PTKR system is that complex quasienergies $\varepsilon=\varepsilon_r \pm \varepsilon_i$ emerge when the strength of the imaginary part of the complex potential exceeds a threshold value, i.e., $\lambda>\lambda_{c}$~\cite{Longhi2017,West2010,Zhao19}. The threshold value $\lambda_c$ is just the exceptional points of the system. It can be proven that the quasieigenstate $|\psi_{\varepsilon}\rangle$ is no longer an eigenstate of the $\cal{PT}$ operator due to the complex quasienergies, thus demonstrating the spontaneous $\mathcal{PT}$-symmetry breaking. An intrinsic quality of the PTKR system is that $\cal{PT}$ symmetry is helpful in protecting the real spectrum of the Floquet operator.

\begin{figure}[htbp]
\begin{center}
\includegraphics[width=9cm]{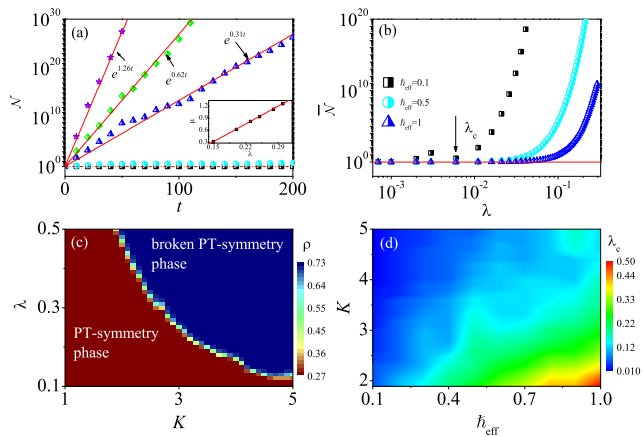}
\caption{(a) Norm $\cal{N}$ versus time for $\lambda=0.01$ (squares), 0.05 (circles), 0.15 (triangles), 0.2 (diamonds), and 0.3 (pentagrams). The parameters are $K=5$ and $\ehbar=1$. Solid lines indicate the exponential increase, i.e., $\mathcal{N}(t)=e^{\mu t}$, while dash-dotted line denotes $\mathcal{N}=1$. Inset: the $\mu$ versus $\lambda$. Solid line indicates the linear increase $\mu \propto\lambda$. (b) Average value $\bar{\cal{N}}$ versus $\lambda$ for $K=5$ with $\ehbar=0.1$ (squares), 0.5 (circles), and 1 (triangles). Solid line denotes $\mathcal{N}=1$. (c) Phase diagram of the spontaneous $\cal{PT}$-symmetry breaking for $\ehbar=1$. One can see clearly a phase boundary $\lambda_c$. (d) The value of $\lambda_c$ in the parameter space $(K,\ehbar)$.
}\label{phasedigm}
\end{center}
\end{figure}

We assume that the initial state is expanded as $|\psi(t_0)\rangle=\sum_{\varepsilon}\rho_{\varepsilon}|\psi_{\varepsilon}\rangle$. Then, after the $n$th kick, the quantum state has the expression $|\psi(t_n)\rangle=U^{t_n}|\psi(t_0)\rangle=\sum_{\varepsilon}\rho_{\varepsilon}e^{-i\varepsilon_r t_n}e^{\varepsilon_i t_n}|\psi_{\varepsilon}\rangle$, whose norm $\mathcal{N}=\langle \psi(t_n)|\psi(t_n)\rangle$ exponentially increases with time due to positive $\varepsilon_i$. We numerically investigate the time evolution of $\cal{N}$ for different $\lambda$. Without loss of generality, we choose a Gaussian wavepacket, i.e., $\psi{(\theta,t_0)}=(\sigma/\pi)^{1/4} \exp (-\sigma \theta^{2}/2)$ with $\sigma=10$ as the initial state in numerical simulations. Figure~\ref{phasedigm}(a) shows that for very small $\lambda$ (e.g., $\lambda=0.01$ and 0.05), the value of $\mathcal{N}$ equals almost to unity with time evolution, which implies that quasienergies are all real. Interestingly, for sufficiently large $\lambda$ (e.g., $\lambda=0.15$), $\mathcal{N}$ increases exponentially with time, i.e., $\mathcal{N}=e^{\mu t}$, and the growth rate $\mu$ increases with the increase of $\lambda$. The non-unitary feature of the Floquet operator, $U_K=\exp[K\lambda\sin(\theta)/\hbar]$, leads to the growth of the norm. A rough estimation of the norm yields a time dependence of the form $\mathcal{N}\propto\exp(K\lambda t/\hbar)$, indicating the relation of the growth rate $\mu\propto \lambda$, which is confirmed by our numerical results [see inset in Figure~\ref{phasedigm}(a)]. We further investigate the long-time average value of the norm, $\bar{\mathcal{N}}=\sum_{n=1}^{N}\mathcal{N}(t_n)/N$, for a wide range of $\lambda$. Figure~\ref{phasedigm}(b) shows that, for a specific $\ehbar$ (e.g., $\ehbar=0.1$), $\bar{\mathcal{N}}$ remains at unity for $\lambda$ smaller than a threshold value $\lambda_c$, beyond which it monotonically increases with $\lambda$. It is reasonable to believe that the threshold value $\lambda_c$ corresponds to the emergence of spontaneous $\cal{PT}$-symmetry breaking.
%%%%

Recently, the long short-term memory (LSTM) network has been exploited to extract the character of time series and thus to predict the phase diagram of quantum diffusion~\cite{Mano21}. Based on the character of the time evolution of $\mathcal{N}$, we conducted supervised training on the LSTM network and used it to evaluate the feature of $\mathcal{N}(t)$, namely, whether $\mathcal{N}(t)=e^{\mu t}$ or not, for different system parameters. This highly effective machine learning method outputs the probability $\rho$ of the time series $\mathcal{N}(t)$ to be exponentially increasing or not, which can predict the phase diagram of spontaneous $\mathcal{PT}$-symmetry breaking. Interestingly, our results show that the $\rho$ increases with the increase of both $K$ and $\lambda$ [see Fig.~\ref{phasedigm}(c)]. We identify two phases in the parameters space $(K,\lambda)$, the boundary of which is clearly visible in Fig.~\ref{phasedigm}(c). We further investigate the $\lambda_c$ for different $K$ and $\ehbar$. Our results demonstrate that the critical parameter $\lambda_c$ increases with the increase of $\ehbar$ and decreases with the increase of $K$ [see Fig.~\ref{phasedigm}(d)]. This behavior is rooted in the fact that the mean spacing level $\Delta$ of the quasienergies of the quantum kicked rotor (QKR) model is proportional to $\hbar/K$~\cite{West2010}. The smaller the $\Delta$ is, the easier it is for the non-Hermitian parameter $\lambda$ to cause the coalescence of two quasienergies, implying the relation $\lambda_c\propto \hbar/K$.

\section{Scaling laws of the OTOCs at the transition to the spontaneous $\cal{PT}$-symmetry breaking}\label{OTCPT}

The OTOCs are defined by $C(t_n) = - \langle [\hat{A}(t_n), \hat{B}]^2\rangle$, with the operators $\hat{A}(t_n)=U^{\dagger}(t_n)AU(t_n)$ and $B$ being evaluated in the Heisenberg picture~\cite{Chen2017,Hashimoto2017,Dora2017,Mata2018,Alavirad2018,Swingle2016,Hafezi2016,Li2017,Garttner2017}. The average, i.e., $\langle \cdots\rangle=\langle \psi(t_0)|\cdots|\psi(t_0)\rangle$, is taken over an initial state $|\psi(t_0)\rangle$~\cite{Heyl2018}. In this work, we consider the case where both $\hat{A}$ and $\hat{B}$ are angular momentum operators, i.e., $C(t) = -\langle [{p}(t), {p}]^2 \rangle$. We use a Gaussian wavepacket, i.e., $\psi{(\theta,t_0)}=(\sigma/\pi)^{1/4} \exp (-\sigma \theta^{2}/2)$ with $\sigma=10$ as the initial state. It is worth noting that, as opposed to static-lattice systems, periodically-driven systems have no thermal states, as the temperature grows to infinity with time evolution~\cite{D'Alessio14}. Thus, there is no need to average over the initially thermal states in the definition of $C(t_n)$ in our system~\cite{WlZhao22,WLZhao21}.

Straightforward derivation yields the equivalence \begin{align}\label{DcompsiOTOC}
C(t_n) =& C_1(t_n) + C_2(t_n)-2\text{Re}\left[C_3(t_n)\right]\;,
\end{align}
where the two-points correlators, namely, the first two terms in right side are defined by
\begin{equation}\label{Firpart}
C_1(t_n) = \langle \psi_R(t_0) |{p}^{2}|\psi_R(t_0)\rangle\;,
\end{equation}
\begin{equation}\label{Secpart}
C_2(t_n) = \langle \varphi_R(t_0)|\varphi_R(t_0)\rangle\;,
\end{equation}
and the four-points correlator is
\begin{equation}\label{Tirdpart}
C_3(t_n) =\langle \psi_R(t_0) |p|\varphi_R(t_0)\rangle\;,
\end{equation}
with $|\psi_R(t_0)\rangle={U}^{\dagger}(t_0,t_n)pU(t_0,t_n)|\psi(t_0)\rangle$ and $|\varphi_R(t_0)\rangle={U}^{\dagger}(t_0,t_n)pU(t_0,t_n)p|\psi(t_0)\rangle$~\cite{Ueda2018}. The symbol $\text{Re}[\cdots]$ denotes the real part of a complex variable.

To obtain the state $|\psi_R(t_0)\rangle$, three steps must be carried out: i) the forward evolution from $t_0$ to $t_n$, i.e. $|\psi(t_n)\rangle=U(t_0,t_n)|\psi(t_0)$, ii) the action of the operator $p$ on the state $|\psi(t_n)\rangle$, i.e. $|\tilde{\psi}(t_n)\rangle= p |\psi(t_n)\rangle$, and iii) the backward evolution from $t_n$ to $t_0$, i.e. $|\psi_R(t_0)\rangle={U}^{\dagger}(t_0,t_n)|\tilde{\psi}(t_n)\rangle$. The expectation value of the square of the momentum can then be calculated using $|\psi_R(t_0)\rangle$ to obtain $C_1(t_n)$ [see Eq.~\eqref{Firpart}]. To numerically simulate $C_2(t)$, the operator $p$ should first be applied to the initial state $|\psi(t_0)\rangle$, yielding the new state $|\varphi(t_0)\rangle = p|\psi(t_0)\rangle$. Then, the forward evolution is conducted, i.e., $|\varphi(t_n)\rangle = U(t_0, t_n)|\varphi(t_0)\rangle$. Subsequently, the action of $p$ is performed on $|\varphi(t_n)\rangle$, obtaining $|\tilde{\varphi}(t_n)\rangle = p|\varphi(t_n)\rangle$, afterwards, the time-reversal is applied to $|\tilde{\varphi}(t_n)\rangle$, resulting in $|\varphi_R(t_0)\rangle = {U}^{\dagger}(t_0, t_n)|\tilde{\varphi}(t_n)\rangle$. Using Eq.~\eqref{Secpart}, $C_2(t_n)$ can then be calculated by evaluating the norm of $|\varphi_R(t_0)\rangle$. Lastly, the term $C_3(t_n)$ [seen in Eq.~\eqref{Tirdpart}] can be determined using the two states $|\psi_R(t_0)\rangle$ and $|\varphi_R(t_0)\rangle$, which is usually complex since they are not identical.

In the $\cal{PT}$-symmetry breaking phase, the norm of the quantum state $\mathcal{N}_{\psi}(t_n)=\langle\psi(t_n) | \psi(t_n) \rangle$ increases exponentially with time regardless of the forward or backward evolution. To address this issue and eliminate its contribution to the OTOCs, we normalize the time-evolved state. For the forward evolution $t_0\rightarrow t_n$ of $|\psi(t_0)\rangle$, we set the norm of the quantum state to be the same as that of the initial state, i.e., $\mathcal{N}_{\psi}(t_j)=\langle\psi(t_0) | \psi(t_0) \rangle$ with $0\leq j \leq n$. The backward evolution starts from the state $|\tilde{\psi}(t_n)\rangle$, whose norm $\mathcal{N}_{\tilde{\psi}}(t_n)=\langle\psi(t_n)|p^2| \psi(t_n) \rangle$ is the mean energy of the state $| \psi(t_n) \rangle$. Thus, it is reasonable to take the norm of the quantum state during the backward evolution $t_n\rightarrow t_0$ to be $\mathcal{N}_{\tilde{\psi}}(t_n)$, i.e., $\mathcal{N}_{\psi_R}(t_j)=\mathcal{N}_{\tilde{\psi}}(t_n)$. In short, the norm of the time-evolved state for both the forward and backward evolution is equal to that of the state it starts from. If the same normalization procedure is applied to the evolution of $|\varphi (t_n)\rangle$, then we will have $\mathcal{N}_{\varphi}(t_j)=\langle \varphi(t_0)|\varphi(t_0)\rangle$ and $\mathcal{N}_{\varphi_R}(t_j)=\langle \tilde{\varphi}(t_n)|\tilde{\varphi}(t_n)\rangle$ ($0\leq j \leq n$) for the forward and time reversal evolutions, respectively.
%%%%
\begin{figure}[t]
\begin{center}
\includegraphics[width=8.0cm]{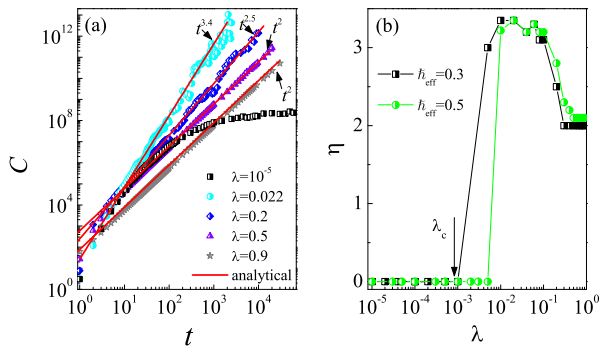}
\caption{(a) Time dependence of $C$ for $K=6$ with $\lambda=10^{-5}$ (squares), $0.022$ (circles), $0.2$ (diamonds), $0.5$ (triangles), and $0.9$ (pentagrams). Solid lines in red denote theoretical prediction in Eqs.~\eqref{SQGLmdc} and \eqref{QGLLmdc}, i.e., $C(t)\propto t^{\eta}$. (b) The $\eta$ versus $\lambda$ for $\ehbar=0.3$ (squares) and 0.5 (circles). Arrow marks the phase transition point $\lambda_c \approx 0.001$ for $\ehbar=0.3$. \label{OTOCLhbar}}
\end{center}
\end{figure}

In order to understand the effects of $\cal{PT}$-symmetry breaking on the dynamics of $C(t_n)$, we numerically investigated the time evolution of $C(t_n)$ for different $\lambda$. Figure~\ref{OTOCLhbar}(a) shows that, for values of $\lambda$ smaller than the phase transition point (e.g., $\lambda=10^{-5}\ll \lambda_c$), the $C(t_n)$ increases gradually up to saturation. Interestingly, for $\lambda$ slightly larger than $\lambda_c$, the $C(t_n)$ increases in a power-law of time, i.e., $C(t_n)\propto t^{\eta}$ with $\eta > 2$ [see  $\lambda = 0.022$ with $\eta =3.4$ in Fig.~\ref{OTOCLhbar}(a)]. We dub this phenomenon as a super-quadratic growth (SQG) of $C(t_n)$. When the value of $\lambda$ is much larger than the phase transition point, i.e., $\lambda\gg\lambda_c$ [e.g., $\lambda = 0.5$ and 0.9 in Fig.2(a)], the quadratic growth (QG) of OTOCs $C(t_n)\propto t^{2}$ emerges. We further investigate the exponent $\eta$ for different $\lambda$. Our results show that $\eta$ is zero for $\lambda<\lambda_c$, increases abruptly to a maximum value greater than two for $\lambda$ slightly larger than $\lambda_c$, and finally saturates to two for sufficiently large $\lambda$ [e.g., see $\ehbar=0.3$ in Fig.~\ref{OTOCLhbar}(b)]. It is evident that the scaling-law of OTOCs reveals the emergence of the spontaneous $\mathcal{PT}$-symmetry breaking and unveils the correlation between information scrambling and the $\mathcal{PT}$-symmetry phase transition.

\section{Theoretical analysis of the dynamics of OTOCs}\label{TheoAnaly}

\subsection{Mechanism of the saturation of $C(t)$ for $\lambda<\lambda_c$}\label{OTCSaturation}

\begin{figure}[b]
\begin{center}
\includegraphics[width=7.5cm]{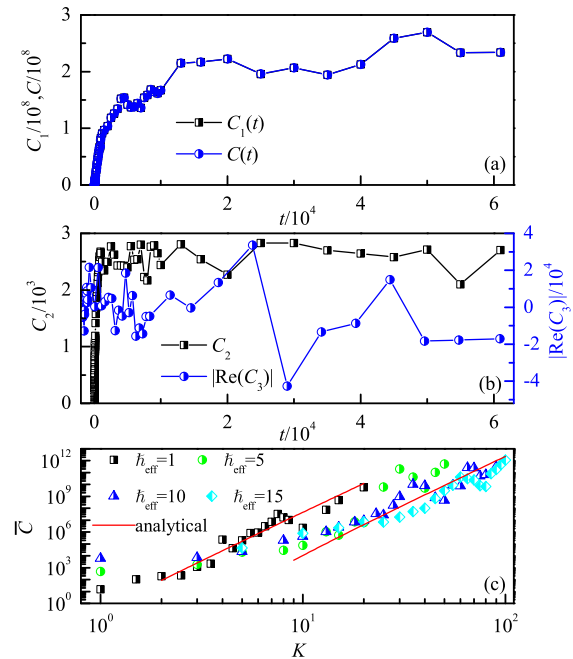}
\caption{(a) Time evolution of $C$ (circles) and $C_1$ (squares). Note that $C$ almost fully overlaps with $C_1$. (b) Dependence of $C_2$ (squares) and $|\textrm{Re}(C_3)|$ (circles) on time. The parameters are $K=6$, $\lambda=10^{-5}$, and $\ehbar=0.3$. (c) The $\bar{C}$ versus $K$ with $\lambda=10^{-5}$ for $\ehbar=1$ (squares), 5 (circles), 10 (triangles), and 15 (diamonds). Solid lines indicates our theoretical prediction in Eq.~\eqref{DLOTOCs}.\label{SLC123}}
\end{center}
\end{figure}

We numerically investigate the time evolution of the three parts of the OTOCs, i.e., $C_1$, $C_2$, and $C_3$ for $\lambda< \lambda_c$. Figure~\ref{SLC123}(a) shows that the time dependence of $C_1$ and $C$ almost overlap, displaying rapid growth up to saturation. Since the real part of $C_3$, i.e., $\textrm{Re}(C_3)$ fluctuates between positive and negative values, we plot the absolute value $|\textrm{Re}(C_3)|$ in Fig.~\ref{SLC123}(b). Both $C_2$ and $|\textrm{Re}[C_3]|$ saturate after a very short time evolution. Importantly, $C_1$ is at least 4 orders of magnitude larger than both $C_2$ and $|\textrm{Re}[C_3]|$, leading to a perfect consistency between $C_1$ and $C$. Consequently, based on Eq.~\eqref{DcompsiOTOC}, we can safely use the approximation
\begin{align}\label{OTOCPP2}
C_{}(t_n) \approx  \langle \psi_R(t_0)|{p}^{2}|\psi_R(t_0)\rangle=\langle p^2(t_0)\rangle_R\mathcal{N}_{\psi_R}(t_0)\;,
\end{align}
where $\langle p^2(t_0)\rangle_R=\langle \psi_R(t_0)|{p}^{2}|\psi_R(t_0)\rangle/\mathcal{N}_{\psi_R}(t_0)$ denotes the exceptional value of energy of the state $|\psi_R(t_0)\rangle$ divided by its norm $\mathcal{N}_{\psi_R}(t_0)=\langle \psi_R(t_0)|\psi_R(t_0)\rangle$.
%%%

The normalization procedure for time reversal yields the equivalence $\mathcal{N}_{\psi_R}(t_0)=\mathcal{N}_{\tilde{\psi}}(t_n)=\langle \psi(t_n)|p^2|\psi(t_n)\rangle$, which shows that the value of $\mathcal{N}_{\psi_R}(t_0)$ is just the mean energy of the state $|\psi(t_n)\rangle$ at the time $t=t_n$. For $\lambda<\lambda_c$, the quasienergies are all real, thus the dynamics of the PTKR is the same as that of the Hermitian QKR. A noteworthy characteristic of the QKR's energy diffusion is the phenomenon of DL, i.e. the mean energy $\langle p^2\rangle$ gradually approaches to saturation level with increasing time due to quantum coherence. It is reasonable to believe that the mechanism of DL suppresses the growth of both $\mathcal{N}_{\psi_R}(t_0)$ and $\langle p^2(t_0)\rangle_R$, and therefore leads to the saturation of $C(t_n)$.

\begin{figure}[htbp]
\begin{center}
\includegraphics[width=8.5cm]{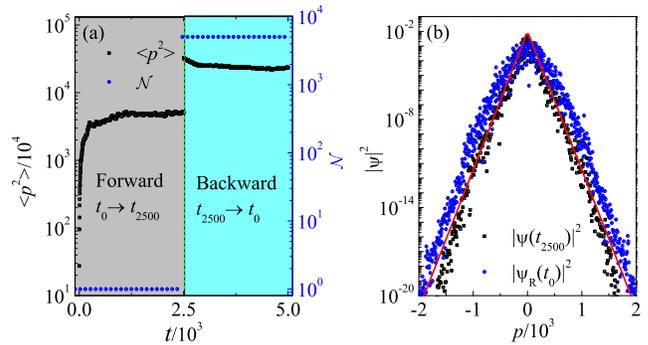}
\caption{(a) Time trace of $\langle p^2 \rangle$ (squares) and $\mathcal{N}$ (circles) during the forward evolution $t_0\rightarrow t_{2500}$, the action of $p$ at the time $t=t_{2500}$, and the time reversal $t_{2500}\rightarrow t_{5000}$. Green dashed line marks $t_n=2500$. (b) Momentum distributions for the state $|\psi(t)\rangle$ (squares) at the time $t_n=2500$ and the state $|\psi_R(t_0)\rangle$ (circles) at the end of time reversal. Solid line indicates the exponentially-localized shape $|\psi(p)|^2\propto e^{-|p|/L}$ with $L\approx 46$. The parameters are $K=6$, $\lambda=10^{-5}$, and $\ehbar=0.3$.\label{SMLTrev}}
\end{center}
\end{figure}

To confirm this conjecture, we consider a specific time, i.e., $t=t_n$, and numerically trace the evolution of $\langle p^2\rangle$ for both the forward ($t<t_n$) and backward ($t>t_n$) evolution. Figure~\ref{SMLTrev}(a) shows that for $t_n=2500$, $\langle p^2\rangle$ increases rapidly to saturation during forward time evolution from $t_0$ to $t_{2500}$, then jumps to a specific value at the start of the time reversal (i.e., at $t=t_{2500}$) before finally saturating for the backward evolution from $t_{2500}$ to $t_0$. This clearly demonstrates the emergence of the DL, which is also reflected by the probability density distribution in momentum space. We compare the momentum distributions at the end of the forward evolution (i.e., $t=t_{2500}$) and the end of time reversal (i.e., $t=t_{0}$) in Fig.~\ref{SMLTrev}(b). One can see that the two quantum states almost overlap with each other, both of which are exponentially localized in momentum space, i.e., $|\psi(t_n)|^2\sim \exp(-|p|/L)$ [see Fig.~\ref{SMLTrev}(b)]. A rough estimation yields $\mathcal{N}_{\psi_R}(t_0)=\langle \psi(t_n)|p^2|\psi(t_n)\rangle\sim L^2$ and $\langle p^2(t_0)\rangle_R\sim \int_{-\infty}^{\infty} p^2\exp(-|p|/L)dp \sim L^2$. Plugging the two relations into Eq.~\eqref{OTOCPP2}, we can immediately get the estimation of the OTOCs, i.e., $C(t_n)\sim L^4$. It is known that the localization length is in a quadratic function of $K$, i.e., $L\propto K^2$~\cite{Izrailev90}, which results in the relation
\begin{equation}\label{DLOTOCs}
C\propto K^8\;.
\end{equation}
This clearly demonstrates that the $C$ is time-independent after the long term evolution, verifying our numerical results in Fig.~\ref{OTOCLhbar}(a) and Fig.~\ref{SLC123}(a).

To provide evidence of our analytical prediction, we investigate the time-averaged value of OTOCs, i.e., $\bar{C}=\sum_{j=1}^{N}C(t_j)/N$, numerically for different $K$. In the numerical simulations, we ensure that $N$ is large enough for the long-term saturation of $C(t)$ to be well quantified by $\bar{C}$. Our numerical results show that for a specific $\hbar$, $\bar{C}$ increases in a power-law of $K$ [see Fig.~\ref{SLC123}(c)], which is well described by our theoretical prediction in Eq.~\eqref{DLOTOCs}. This is a strong indication of the validity of our analytical analysis. Our findings of the dependence of the OTOCs on the kick strength provide an opportunity to control the operator growth with an external driven potential.
%%%%%

The discontinuous jump in the mean square momentum, $\langle p^2\rangle$, at $t=t_{2500}$, the beginning of time reversal, is due to the action of the operator $p$ on the quantum state $|\psi(t_{2500})\rangle$. This action generates the quantum state $|\tilde{\psi}(t_{2500})\rangle=p|\psi(t_{2500})\rangle$, for which the mean value is given by $\langle\tilde{\psi}(t_{2500})|p^2|\tilde{\psi}(t_{2500})\rangle=\langle\psi(t_{2500})|p^4|\psi(t_{2500})\rangle$. The exponentially-localized shape of the quantum state, $|\psi(t_{2500})|^2\sim \exp(-|p|/L)$ [see Fig.~\ref{SMLTrev}(b)], allows us to obtain the expectation values $\langle \psi(t_{2500})|p^2|\psi (t_{2500})\rangle \sim L^2\ehbar^2$ and $\langle \tilde{\psi}(t_{2500})|p^2|\tilde{\psi}(t_{2500})\rangle= \langle \psi(t_{2500})|p^4|\psi(t_{2500})\rangle\sim L^4\ehbar^4$, which quantitatively explains the discontinuous increase in the mean energy from $L^2\ehbar^2$ to $L^4\ehbar^4$.

\subsection{Mechanism of the SQG of $C(t)$ for $\lambda \gtrsim \lambda_c$}

\begin{figure}[htbp]
\begin{center}
\includegraphics[width=7.0cm]{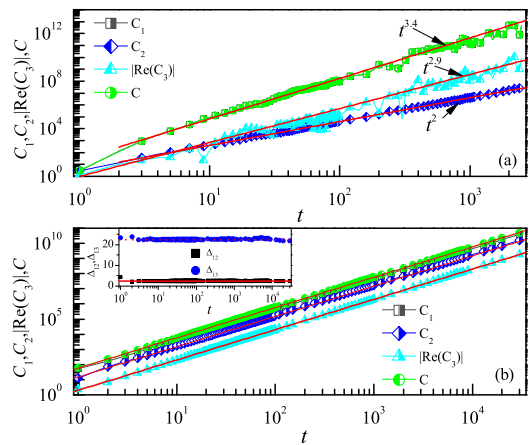}
\caption{Time evolution of $C_1$ (squares), $C_2$ (diamonds), $|\textrm{Re}(C_3)|$ (triangles), and $C$ (circles) for $\lambda=0.022$ (a) and 0.9 (b). In (a): Red lines indicates the power-law fitting. In (b): Red lines indicate the quadratic function $C_1 \approx 44t^2$, $C_2 \approx 17t^2$, and $|\textrm{Re}(C_3)| \approx 2t^2$. Inset: The $\Delta_{12}$ (squares) and $\Delta_{13}$ (circles) versus time. Solid (dashed-dotted) line denotes $\Delta_{12}\approx2.5$ ($\Delta_{13}\approx22$). Other parameters are same as Fig.~\ref{SMLTrev}. \label{OTCLmdc}}
\end{center}
\end{figure}

Figure~\ref{OTCLmdc}(a) shows the time evolution of $C_1$, $C_2$, $|\textrm{Re}(C_3)|$, and $C$ for $\lambda$ just larger than the $\cal{PT}$-symmetry phase transition point $\lambda \gtrsim \lambda_c$. It is clear that the time dependence of $C_1$ corresponds perfectly to that of $C$, both of which increase following the SQG $C \propto t^{3.4}$. The time evolution of $C_2$ displays the QG, i.e., $C_2(t)\propto t^2$, while $|\textrm{Re}(C_3)|$ follows the SQG $|\textrm{Re}(C_3)|\sim t^{2.9}$. In addition, one can see that $C_1$ is larger than both $C_2$ and $|\textrm{Re}(C_3)|$ by approximately four orders of magnitude. Therefore, it is sufficient to analyze the time evolution of the term $C_1$ to uncover the mechanism of the SQG of $C$.
%%%%%

\begin{figure}[htbp]
\begin{center}
\includegraphics[width=8.0cm]{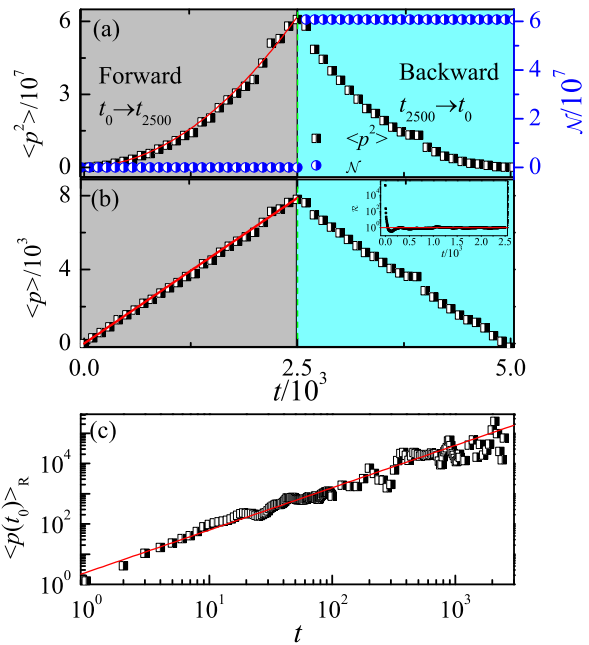}
\caption{Top two panels: Time trace of $\langle p^2 \rangle$ (squares) in (a), $\mathcal{N}$ (circles) in (a), and  $\langle p \rangle$ in (b) during the forward evolution $t_0 \rightarrow t_{2500}$, the action of $p$ at the time $t=t_{2500}$, and the time reversal $t_{2500} \rightarrow t_0$. Green dashed lines makes $t=t_{2500}$. In (a): Red line indicates the quadratic function $\langle p^2\rangle = \gamma^2t^2$ with $\gamma\approx 3.2$. In (b): Red line indicates the linear growth $\langle p\rangle = \gamma t$. Inset: $\cal{R}$ versus time. Red line marks $\mathcal{R}=1$. In (c): Time dependence of $\langle p^2(t_0)\rangle_R$. Red line indicates the power-law fitting $\langle p^2(t_0)\rangle_R \propto t^{1.4}$. The parameter is $\lambda=0.022$. Other parameters are same as Fig.~\ref{SMLTrev}. \label{TimRLmdc}}
\end{center}
\end{figure}

Since the value of $C(t)$ at a specific time $t=t_n$ is dependent on both the mean energy $\langle p^2(t_0)\rangle_R$ and the norm $\mathcal{N}_{\psi_R}(t_0)$ of the state $|\psi_R(t_0)\rangle$ at the end of time reversal [see Eq.~\eqref{OTOCPP2}], we numerically calculate the forward and backward time evolution of $\langle p^2\rangle$, $\langle p\rangle$, and $\mathcal{N}$ with a fixed $t_n$ (e.g., $t_n=2500$ in Fig.~\ref{TimRLmdc}). Figure~\ref{TimRLmdc}(a) demonstrates that the mean energy diffuses ballistically with time $\langle p^2\rangle\approx \gamma^2 t^2$ during the forward evolution $t_0 \rightarrow t_{2500}$, and it displays the intrinsic time reversal during $t_{2500}\rightarrow t_0$. Meanwhile, the mean momentum $\langle p\rangle$ linearly increases for $t_0< t< t_{2500}$, and linearly decays for $t_{2500} \rightarrow t_{0}$ [see Fig.~\ref{TimRLmdc}(b)]. Moreover, Fig.~\ref{TimRLmdc}(a) reveals that the norm remains unity, i.e., $\mathcal{N}(t)=1$, during the forward evolution and equals the mean energy at the time $t=t_{2500}$, i.e., $\mathcal{N}_{\psi_R}(t_j)=\langle p^2(t_{2500})\rangle$ during the time reversal. Taking the ballistic diffusion of energy into account, the following equivalence can be derived
\begin{equation}\label{NormETR}
\mathcal{N}_{\psi_R}(t_0)=\gamma^2 t_n^2\;.
\end{equation}

In order to measure the degree of time reversal for a fixed $t_n$, we define the ratio of mean energy between forward ($t<t_n$) and backward ($t>t_n$) time evolution as
\begin{equation}\label{DGETRever}
\mathcal{R}(t_j)=\frac{\langle p^2(2t_n-t_j)\rangle_R}{\langle p^2(t_j)\rangle}\;, \end{equation}
where $\langle p^2(2t_n-t_j)\rangle_R$ and $\langle p^2(t_j)\rangle$ ($0\leq j\leq n$) denote the mean square of momentum for the forward evolution and time reversal, respectively. The inset in Fig.~\ref{TimRLmdc}(b) shows that $\mathcal{R}$ is very large (i.e., $\mathcal{R}\gtrsim 10^4$ for $t=t_0$) and approaches almost one with time evolution. This reveals that the mean energy at the end of time reversal is much greater than that at the initial time, i.e., $\langle p^2(t_0)\rangle_R\gg \langle p^2(t_0)\rangle$. We further investigate the time evolution of $\langle p^2(t_0)\rangle_R$, and find [see Fig.~\ref{TimRLmdc}(c)] that the $\langle p^2(t_0)\rangle_R$ increases in the power-law of time
\begin{equation}\label{EnergyETR}
\langle p^2(t_0)\rangle_R\propto t^{1.4}\;.
\end{equation}
Substituting Eqs.~\eqref{NormETR} and~\eqref{EnergyETR} into Eq.~\eqref{OTOCPP2} yields the SQG of OTOCs
\begin{equation}\label{SQGLmdc}
C(t)\propto t^{\eta}\quad \text{with}\quad \eta\approx 3.4 \;.
\end{equation}
%%%%%

\begin{figure}[ht]
\begin{center}
\includegraphics[width=8.5cm]{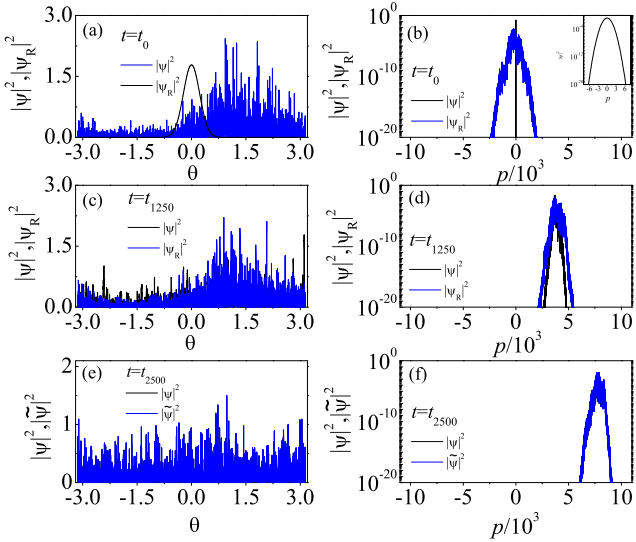}
\caption{Probability density distributions in real (left panels) and momentum (right) space for $\lambda=0.022$. (a)-(d) Black and blue lines separately correspond to the states at forward $|\psi(t)\rangle$ and backward $|\psi_R(t)\rangle$ evolution with $t=t_0$ (top panels) and $t=t_{1250}$ (middle panels). Inset in (b) shows a magnified view of the momentum distribution for the initial Gaussian wavepacket around $p=0$. Bottom panels: Probability density distributions in real (e) and momentum (f) space at the time $t=t_{2500}$. Black and blue lines indicate the state $|\psi(t_{2500})\rangle$ and $|\tilde{\psi}(t_{2500})\rangle=p|\psi(t_{2500})\rangle$, respectively. Other parameters are same as Fig.~\ref{SMLTrev}.\label{lam0.022states}}
\end{center}
\end{figure}

Figure~\ref{lam0.022states} shows the probability density distribution of the state at forward $|\psi(t_j)\rangle$ and backward $|\psi_R(t_j)\rangle$ evolution in both the real and momentum space. The initial state is a Gaussian wavepacket  $\psi{(\theta,t_0)}=(\sigma/\pi)^{1/4} \exp (-\sigma \theta^{2}/2)$ centered at $\theta=0$ and $p=0$ [see Figs.~\ref{lam0.022states}(a) and (b)]. Interestingly, one can observe that the quantum state is mainly distributed in the region $0<\theta<\pi$ for both the forward and backward evolution. This is due to the fact that the action of the Floquet operator of the kicking term $U_K(\theta)=\exp[K\lambda\sin(\theta)/\ehbar]\exp[-iK\cos(\theta)/\ehbar]$ on a quantum state, i.e., $U_K(\theta)\psi(\theta)$ helps to amplify the state within the region $0<\theta<\pi$ as $K\lambda \sin(\theta)>0$. Assuming that the real part of the kicking potential provides the driven force $F=K\sin(\theta)$, the PTKR experiences a positive magnitude force $F>0$ during the forward evolution, thus the momentum grows with time, as shown in Fig.~\ref{TimRLmdc}(b). For the time reversal, the sign of kick strength $K$ flips, i.e., $K \rightarrow -K$, so the mean momentum decreases with time evolution. Our conjecture is supported by the numerical results of momentum distributions. Figures~\ref{lam0.022states}(b), (d) and (f) show that the wavepacket, like a soliton, moves to the positive direction in momentum space for $t_0 \rightarrow t_{2500}$, resulting in $\langle p\rangle=\gamma t$ [in Fig.~\ref{TimRLmdc}(b)], and moves back to the opposite direction for $t_{2500} \rightarrow t_{0}$. In addition, the width of the wavepacket in momentum space is so narrow that one can safely use the approximation $\langle p^2 \rangle \sim (\langle p\rangle)^2$, which is verified by our numerical results in Figs.~\ref{TimRLmdc}(a).
%%%%%%

\subsection{Mechanism of the QG of $C(t)$ for $\lambda \gg \lambda_c$}

We numerically investigate the time evolution of $C_1$, $C_2$, $C_3$, and $C$ for $\lambda\gg \lambda_c$. As shown in Fig.~\ref{OTCLmdc}(b), all of them increase in the way of QG (i.e., $\propto t^2$). We use the ratios $\Delta_{12}=C_1/C_2$ and $\Delta_{13}=C_1/|\textrm{Re}[C_3]|$ to quantify the differences among $C_1$, $C_2$, and $|\textrm{Re}[C_3]|$. Our investigation show that both of them are larger than one, specifically $\Delta_{12}\approx 2.5$ and $\Delta_{13}\approx 22$ [see the inset in Fig.~\ref{OTCLmdc}]. This suggests that $C_1$ contributes mainly to the $C$, which is verified by the good agreement between $C_1$ and $C$ [see Fig.~\ref{OTCLmdc}(b)]. To reveal the mechanism of the QG of $C$, we proceed to analyze the time evolution of $C_1$ by thoroughly investigating both the forward and backward evolution of the mean values $\langle p^2\rangle$, $\langle p\rangle$, and the norm $\cal{N}$ for a given $t_n$.

\begin{figure}[b]
\begin{center}
\includegraphics[width=8.0cm]{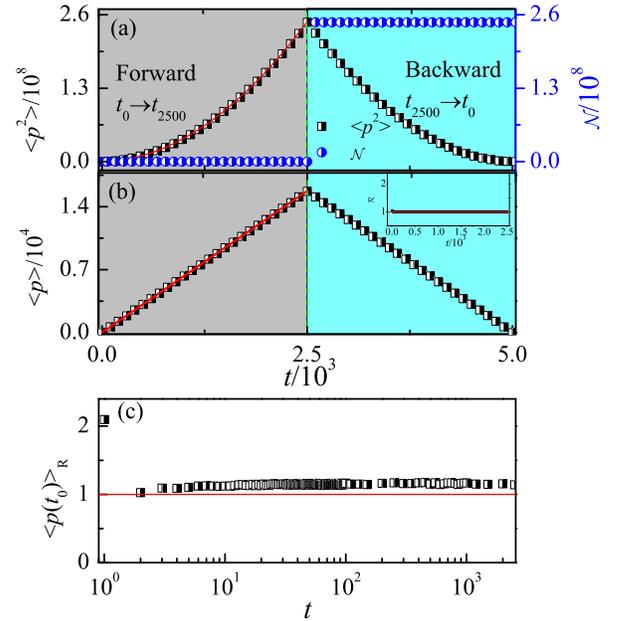}
\caption{Same as in Fig.~\ref{TimRLmdc} but for $\lambda=0.9$. In (a): Red line indicates the quadratic function $\langle p^2\rangle = \gamma^2t^2$ with $\gamma\approx 6.3$. In (b): Red line denotes the linear growth $\langle p\rangle = \gamma t$. In (c): Red line indicates $\langle p(t_0)\rangle_R \approx 1$.\label{TimRLGLmd}}
\end{center}
\end{figure}

Figure~\ref{TimRLGLmd}(a) shows that the mean energy exhibits ballistic diffusion $\langle p^2\rangle\approx \gamma^2 t^2$, during the forward evolution from $t_0$ to $t_n$, and decays as the inverse of a quadratic function, with $\langle p^2\rangle\propto t^{-2}$, during the backward evolution from $t_n$ to $t_0$. This decay is symmetric with respect to the $\langle p^2\rangle$ of $t<t_n$. The dynamics of the mean momentum also exhibits perfect time reversal, namely it linearly increases as $\langle p\rangle =\gamma t$ during $t_0\rightarrow t_{2500}$ and decreases linearly during $t_{2500}\rightarrow t_0$. The ratio $\mathcal{R}$ remains close to one throughout the time evolution, except at the end, i.e., $\mathcal{R}(t=2500)\approx 2.5$ [see the inset in Fig.~\ref{TimRLGLmd}(b)], providing a clear evidence of time reversal.
For the forward evolution from $t_0$ to $t_{2500}$, the norm $\mathcal{N}(t_j)$ is equal to unity, while for the interval $t_{2500}\rightarrow t_{0}$, it is equal to the value of $\langle p^2(t_{2500})\rangle$, i.e., $\mathcal{N}_{\psi_R}(t_j)=\langle p^2(t_{2500})\rangle$ [see Fig.~\ref{TimRLGLmd}(a)]. By utilizing the ballistic diffusion of mean energy, we establish the relationship $\mathcal{N}_{\psi_R}(t_0)\approx \mathcal{N}_{\psi_R}(t_n)\approx \gamma^2t_n^2$, where $t_n$ is an arbitrary time. We further evaluate the behavior of $\langle p^2(t_0)\rangle_R$ for different $t_n$. As shown in Fig.~\ref{TimRLGLmd}(c), the $\langle p^2(t_0)\rangle_R$ remains almost constant at a value of one, indicating that it is independent of time. Plugging in the values of $\mathcal{N}_{\psi_R}(t_0)$ and $\langle p^2(t_0)\rangle_R$ into Eq.~\eqref{OTOCPP2}, we obtain the QG of OTOCs
\begin{equation}\label{QGLLmdc}
C(t)\approx \gamma^{2} t^{\eta}\quad \text{with}\quad \eta=2.
\end{equation}
%%%%%

\begin{figure}[t]
\begin{center}
\includegraphics[width=8.5cm]{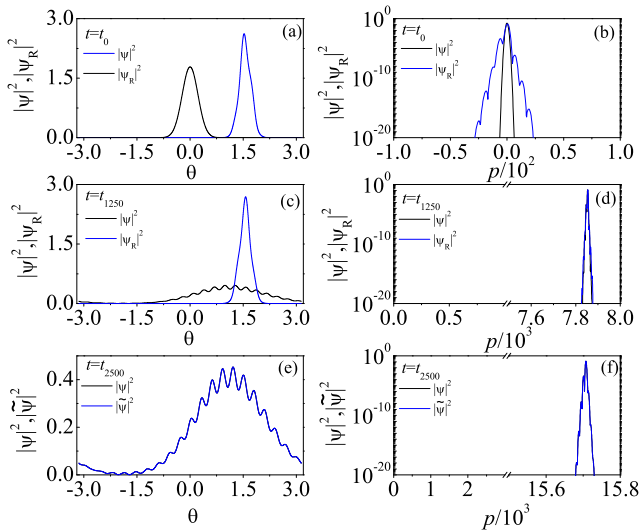}
\caption{Same as in Fig.~\ref{lam0.022states} but for $\lambda=0.9$.\label{lamd09states}}
\end{center}
\end{figure}

The time reversal of a wavepacket's dynamics is clearly seen in the evolution of its momentum distributions. For the forward time evolution, the quantum state is localized at the point $\theta_c=\pi/2$ [seen in Figs.~\ref{lamd09states}(a), (c) and (e)], which is the result of the localization-effect of the imaginary part of the Floquet operator $U_K(\theta)$. With the wavepacket mimicking a classical particle, it experiences a kicking force of magnitude $F=K \sin(\theta_c)=K$, resulting in a constant acceleration of momentum $\Delta p=K$, which is reflected in the linear growth of momentum. This phenomenon of the directed current is also seen in the propagation of momentum distributions in Figs.~\ref{lamd09states}(b), (d) and (f), where a soliton can be observed moving unidirectionally towards the positive direction in momentum space.

During the backward evolution, the wavepacket of the real space remains centered at $\theta_c=\pi/2$, with a width much smaller than the corresponding state at the time of forward evolution [see Figs.~\ref{lamd09states}(a), (c) and (e)]. As the particle is exposed to the kicking force with $F=-K$ during time reversal, its momentum decreases linearly in time, which is also reflected in the propagation of the wavepackets in momentum space [Figs.~\ref{lamd09states}(b), (d) and (f)]. It is evident that the $|\psi(t_n)|^2$ is in perfect overlap with the $|\psi_R(t_n)|^2$, apart from the initial state [see Figs.~\ref{lamd09states}(b)]. The width of $|\psi_R(t_0)|^2$ is considerably larger than that of $|\psi(t_0)|^2$, leading to the ratio of energy being larger than one, i.e., $\mathcal{R}=\langle p^2(t_{5000})\rangle_R/\langle p^2(t_0)\rangle\approx 2.5$ [see the inset in Fig.~\ref{TimRLGLmd}(b)].
%%%%%%%

\section{Conclusion and discussion}\label{Sum}

In this work, we investigate the dynamics of OTOCs in a PTKR model and achieve its scaling laws in different phases of $\cal{PT}$ symmetry. We use the time series of the norm to train a LSTM, which enables us to extract a clear phase diagram of $\cal{PT}$-symmetry breaking, with a phase boundary at $\lambda_c$. For $\lambda<\lambda_c$, we find that the DL of energy diffusion suppresses the growth of OTOCs, and prove analytically the dependence of OTOCs on the kicking strength, i.e., $C\propto K^8$. At the vicinity of the phase transition points, i.e., $\lambda \gtrsim \lambda_c$, we observe a SQG of OTOCs, i.e., $C(t)\propto t^{\eta}$ with an exponent $\eta>2$. Interestingly, a QG of OTOCs, i.e., $C(t)\propto t^2$ emerges for $\lambda\gg \lambda_c$. We elucidate the mechanisms of both the SQG and QG by analyzing the time-reversed wavepacket's dynamics. Our results demonstrate that the spontaneous $\cal{PT}$-symmetry breaking profoundly affects the dynamics of OTOCs, providing an unprecedented opportunity for diagnosing the spontaneous $\cal{PT}$-symmetry breaking with OTOCs.

In recent years, the OTOCs have been widely used to investigate the operator growth in quantum mapping systems~\cite{Moudgalya19}, the information scrambling in spin chains~\cite{TianciZhou20}, and the quantum thermalization in many-body chaotic systems~\cite{KenXuanWei19}. Theoretical studies have demonstrated that the QKR model is mathematically equivalent to the kicked Heisenberg spin XXZ chain~\cite{Boness10}, indicating a connection between the magnon dynamics and quantum diffusion of chaotic systems. Our findings therefore bridge the gap between the information scrambling in condensed matter physics and the operator growth in quantum chaotic systems. This also paves the way for the experimental observation of OTOCs dynamics in chaotic systems using spin chain platforms.

\section*{ACKNOWLEDGMENTS}
Wen-Lei Zhao is supported by the National Natural Science Foundation of China (Grant Nos. 12065009), the Natural Science Foundation of Jiangxi province (Grant Nos. 20224ACB201006 and 20224BAB201023) and the Science and Technology Planning Project of Ganzhou City (Grant No. 202101095077). Jie Liu is supported by the NSAF (Contract No. U1930403).

\end{CJK*}  %% end Chinese, Japanese, and Korea languages environment

\end{document}